 \documentclass[twoside,reqno,11pt]{fcaa}

\usepackage{graphicx}
\usepackage{epsfig}
\usepackage{amsthm}
\usepackage{amsmath}
\usepackage{latexsym}
\usepackage{amsfonts}
\usepackage{amssymb}

\newtheoremstyle{theorem}
  {15pt}          
  {15pt}  
  {\sl}  
  {\parindent}
  {\sc}  
  {. }   
  { }    
  {}     
\theoremstyle{theorem}

\newtheoremstyle{defi}
  {15pt}          
  {15pt}  
  {\rm}  
  {\parindent}     
  {\sc}  
  {. }    
  { }    
  {}     
\theoremstyle{defi}

 \def\proofname{\indent\rm P\ r\ o\ o\ f.}
 
 \def\qed{\hfill$\Box$} 

 \usepackage{hyperref} 

\usepackage{cite}

 \def\theequation{\arabic{section}.\arabic{equation}}

 \def\themycopyrightfootnote{\vspace*{3pt}
 \copyright \, Year\,  Diogenes  Co., Sofia
   \par  \noindent pp. xxx--xxx
   \hfill  \vspace*{-36pt}
  }

 \title[GENERALIZED DIFFUSION EQUATION WITH NONLOCALITY \dots]{Generalized diffusion equation with nonlocality of space-time: analytical and numerical analysis}
 \author[P. Kostrobij, M. Tokarchuk, B. Markovych, I. Ryzha]{Petro Kostrobij $^1$, Mykhailo Tokarchuk $^{2,1}$, Bogdan Markovych $^1$, Iryna Ryzha $^1$}

\newcommand{\rd}{\mathrm{d}}
\newcommand{\ri}{\mathrm{i}}
\newcommand{\re}{\mathrm{Re}\,}
\newcommand{\im}{\mathrm{Im}\,}
\newcommand{\ee}{\mathrm{e}}
\newcommand{\DDD}[3]{\raisebox{-.6ex}{\scriptsize #1}\mathrm{D}_{#2}^{#3}}

\begin{document}
 \newcommand{\doi}[1]{{\texttt{#1}}}

 \sloppy

 \vbox to 2.5cm { \vfill }


 \bigskip \medskip

 \begin{abstract}
 Based on the non-Markov diffusion equation taking into account the spatial fractality and modeling for the generalized coefficient of particle diffusion $D^{\alpha\alpha'}(\mathbf{r},\mathbf{r}';t,t')=W(t,t')\overline{D}^{\alpha\alpha'}(\mathbf{r},\mathbf{r}')$  using fractional calculus the generalized Cattaneo--Maxwell--type diffusion equation in fractional time and space derivatives has been obtained.
 In the case of a constant diffusion coefficient,
 analytical and numerical studies of the frequency spectrum for the Cattaneo--Maxwell diffusion equation in fractional time and space derivatives have been performed. Numerical calculations of the phase and group velocities with change of values of characteristic relaxation time, diffusion coefficient and indexes of temporal $\xi$ and spatial $\alpha$ fractality have been carried out.
 \medskip


 \smallskip


 \end{abstract}

 \maketitle

 \vspace*{-16pt}


\section{Introduction}\label{Sec:1}
\setcounter{section}{1}
\setcounter{equation}{0}

 Fractional integrals and derivatives~\cite{Oldham2006,Samko1993,Podlubny1998,Mandelbrot1982,Uchaikin2008500} are actively used in researches of anomalous diffusion in porous media~\cite{Uchaikin2008500,Sahimi1998213,Korosak20071,Metzler20001,Hilfer2000,Bisquert20002287,Bisquert2001112,Kosztolowicz2009055004,Pyanylo201484,Zhokh20187176,Zhokh201735,Zhokh2017124704},
 disordered systems~\cite{Scher19752455,Berkowitz19985858,Bouchaud1990127,Nigmatullin1984739,Nigmatullin1984389,Nigmatullin1986425,Nigmatullin1992242,Nigmatullin199787,Nigmatullin2009014001,Khamzin20121604,Popov20121,Grygorchak2015e,Kostrobij2015154,Kostrobij20184099},
 plasma physics~\cite{Balescu19954807,Tribeche2011103702,Gong2012023704,Carreras20015096,Tarasov2005082106,Tarasov2006052107},
 turbulent~\cite{Monin1955256,Klimontovich2002,Zaslavsky2002461},
 kinetic and reaction-diffusion processes~\cite{Tarasov2010,Zaslavsky2002461,Zaslavsky1994110,Saichev1997753,Zaslavsky2004128,Nigmatullin2006282,Chechkin200278,
 Gafiychuk2007055201,Datsko2018237,Kosztolowicz2008066103,Shkilev20131066,Baron2019052124},
 in quantum mechanics~\cite{Laskin2000780,Laskin20003135,Laskin2000298,Laskin2002056108,Naber20043339}, viscoelastic~\cite{Magin20101586,Keshavarz2017663,Bonfanti20206002,Makris2020} and  biological systems~\cite{Hobbie2007,Jeon2012188103,Hofling2013046602},
 etc.~\cite{Uchaikin2008500,Uchaikin20131074,Szymanski2009038102}.

 Experimental data on different processes of anomalous diffusion show that
 not only the distribution law,
 but also form of diffusion package is significantly different from the normal diffusion~\cite{Uchaikin2008500,Bouchaud1990127,Zaslavsky2002461,Uchaikin20131074}.
 Approaches with variable diffusion coefficients~\cite{O'Shaughnessy1985455},
 on the basis of degree correlations of fractional order~\cite{Mandelbrot1982},
 fractional derivatives~\cite{Monin1955256,Klimontovich2002,Zaslavsky2002461},
 the generalized Fokker-Planck equation~\cite{Metzler20001,Zaslavsky2002461,Metzler1999431},
 generalizations of statistical mechanics (extensive and non-extensive)
 based on the Tsallis~\cite{Essex2000299,Tsallis2001,Gell-Mann2004} and
 Renyi~\cite{Essex2000299,Vasconcellos20064821} entropy,
 and others were developed to describe anomalous diffusion in different physical and chemical systems.
 Conducted researches show that
 mathematical basis of anomalous diffusion is equation with fractional derivatives~\cite{Uchaikin2008500,Zaslavsky2002461}.
 In particular,
 during the study of three-dimensional models of anomalous diffusion~\cite{Uchaikin2008500,Uchaikin20131074,Uchaikin2003810},
 basic equations of anomalous diffusion are derived
 from the general principles of the stochastic theory of random processes
 based on the Chapman-Kolmogorov integral equations for transition probabilities.
 Solutions of these equations form a new class of distributions,
 which are called fractional stable distributions.
 These distributions are solutions of partial differential equations of fractional order.
 These equations are generalization of usual diffusion equation to the case of anomalous diffusion.
 A partial case of the fractional stable distributions is the Gaussian distribution,
 which corresponds to the normal diffusion.
 It is important to note that obtained equations for anomalous diffusion with fractional derivatives
 contain diffusion coefficient,
 which is a constant in time and space.
 On the other hand,
 diffusion coefficients are related to time correlation functions
 (the Green-Kubo relations),
 which contain diffusion transfer mechanisms from the perspective of nonequilibrium statistical mechanics.

 Currently,
 together with phenomenological approaches for constructing of the Fokker-Planck equation,
 the diffusion equation and its generalization --- the Cattaneo equation with fractional derivatives,
 there are two methods of constructing such equations,
 namely,
 (1)~probabilistic method,
 which is based on the Chapman-Kolmogorov equation in the stochastic theory of random processes~\cite{Uchaikin2008500,Zaslavsky2002461,Stanislavsky2004418},
 and (2)~statistical method,
 which is based  on the method of projection operators (memory functions) in the works
 ~\cite{Nigmatullin1984739,Nigmatullin1984389,Nigmatullin1986425,Nigmatullin1992242,Nigmatullin199787,
 Nigmatullin2009014001,Nigmatullin2006282,Khamzin20121604},
 and
 on the Liouville equation with fractional
 derivatives~\cite{Tarasov2004123,Tarasov200517,Tarasov2005011102,Tarasov2006033108,Tarasov2006341,Tarasov2013663,Tarasov20082984,Tarasov2005286,Tarasov2007237,Tarasov2007163,Tarasov2009179,Tarasov20121719,Tarasov2013102110,Tarasov2010}.
 In particular,
 by using this method,
 the BBGKY hierarchy equations with fractional derivatives~\cite{Tarasov200517,Tarasov2005011102,Tarasov2007237},
 transport equation,
 diffusion equation,
 and the Heisenberg equation with fractional derivatives~\cite{Tarasov2006341,Tarasov2013663,Tarasov20082984} are obtained.
 This approach is formulated for non-Hamiltonian systems.
 If the Helmholtz conditions for coordinate and momentum derivatives of fields of velocities and forces,
 which act on particles,
 are fulfilled,
 the Hamiltonian systems with the time-reversible Liouville equation with fractional derivatives are obtained
 from non-Hamiltonian systems.
 In Ref.~\cite{Kobelev20000002002},
 time-irreversible equations of motion of Hamilton and Liouville
 for dynamic of classical particles in space with multifractal time are offered.
 By using the definition of fractional derivative and the Riemann-Liouville integral,
 the time-irreversible Liouville equation with fractional derivatives
 (where the time is given on multifractal sets with fractional dimensions) is obtained.
 In Refs.~\cite{Kobelev2000194,Kobelev2002580},
 kinetic equations for systems with fractal structure
 (in particular, for description of diffusion processes in space of coordinates and momenta) are obtained
 within the Klimontovich approach.
 A similar approach for constructing of time fractional generalization
 for the Liouville equation and the Zwanzig equation (within projection formalism) is proposed in Ref.~\cite{Lukashchuk2013740}.

 An actual problem for description of nonequilibrium processes in complex systems is construction of generalized diffusion and wave equations~\cite{Sandev2018015201,Sandev2019} using fractional integrals and derivatives.
 The dispersion of heat waves in a dissipative environment using the Cattaneo--Maxwell heat diffusion equation with fractional derivatives has been investigated in Ref.~\cite{Giusti2018013506}.
 On the basis of this equation,
 the frequency spectrum, phase and group velocities of propagation of heat waves in a dissipative environment have been investigated.

 It is important to note that,
 for the first time,
 in Refs.~\cite{Nigmatullin1984739,Nigmatullin1984389,Nigmatullin1986425},
 Nigmatullin received diffusion equation with the fractional time derivatives for the mean spin density~\cite{Nigmatullin1984739},
 the mean polarization~\cite{Nigmatullin1984389},
 and the charge carrier concentration~\cite{Nigmatullin1986425}.
 In Ref.~\cite{Nigmatullin1992242},
 justification of equations with fractional derivatives is given,
 and the time irreversible Liouville equation with the fractional time derivative is provided.
 Within this approach,
 some important results,
 including microscopic model of a non-Debye dielectric relaxation,
 which generalizes the Cole-Cole law~\cite{Khamzin20121604} and the Cole-Davidson law~\cite{Nigmatullin199787}, are obtained.
 In Ref.~\cite{Popov20121},
 by using the fractal nature of transport processes of charge carriers,
 low-frequency behavior of conductivity is studied with taking into account polarization effects of electrode.
 Results of this investigation are in good agreement with experimental data.

 In our works~\cite{Kostrobij201963,Kostrobij2016093301,Kostrobij2016163,Glushak201857,Kostrobij201875,Kostrobij201958,Grygorchak2015e,
 Kostrobij2015154,Grygorchak2017185501,Kostrobij20184099,Kostrobij2019289} a statistical approach to obtain generalized spatiotemporal nonlocal transfer equations was developed by using the Zubarev nonequilibrium statistical operator (NSO) method~\cite{Zubarev19811509,Zubarev20021and2,Markiv2011785} and the Liouville equation with fractional derivatives~\cite{Tarasov2010,Tarasov2004123}.
 In particular,
 the generalized diffusion equations of Cattaneo~\cite{Kostrobij2016163,Kostrobij201875,Kostrobij201963},
 Cattaneo--Maxwell~\cite{Kostrobij201958} and electrodiffusion~\cite{Grygorchak2015e,Kostrobij2015154,Grygorchak2017185501,Kostrobij20184099},
 kinetic equations~\cite{Kostrobij2019289} with spatiotemporal fractional derivatives were obtained.

 In the second section,
 based on the statistical approach within the Gibbs statistics~\cite{Kostrobij202023003},
 we have obtained a generalized diffusion equation  with fractional derivatives for the nonequilibrium average value of the number density of particles.
 This equation is nonlocal in space and time.

 By modeling the temporal and spatial dependence of memory function using fractional calculus~\cite{Oldham2006,Samko1993,Podlubny1998,Mandelbrot1982,Uchaikin2008500}, the generalized Cattaneo--Maxwell--type diffusion equation has been obtained and analyzed.

 In the third section,
 within the Gibbs statistics and approximation of constant diffusion coefficient,
 the frequency spectrum of the Cattaneo--Maxwell--type diffusion equation for the nonequilibrium average value of the number density of particles has been obtained.
 The frequency spectrum,
 phase and group velocities have been calculated,
 depending on order of the fractional derivative,
 characteristic relaxation time and value of the diffusion coefficient.


\section{Generalized diffusion equations with fractional derivatives}
\setcounter{section}{2}
\setcounter{equation}{0}

 To describe the diffusion processes of particle in heterogeneous environments with fractal structure,
 one of main parameters of the reduced description is the nonequilibrium density of particle numbers
 $n (\mathbf{r};t)=\langle \hat{n} (\mathbf{r})\rangle _{\alpha }^{t}$,
 where $\hat{n} (\mathbf{r})=\sum _{j=1}^{N} \delta (\mathbf{r}-\mathbf{r}_{j} )$ is the microscopic density of the particle.
 The corresponding generalized diffusion equation for $n (\mathbf{r}; t)$ can be obtained on the base of approach~\cite{Kostrobij2016093301},
 by using the Zubarev nonequilibrium statistical operator method
 within the Gibbs statistics for solution of the Liouville equations with fractional derivatives,\vspace{-1mm}
 \begin{equation} \label{01}
    \frac{\partial }{\partial t} \left\langle \hat{n} (\mathbf{r})\right\rangle _{\alpha }^{t} =
    \mathrm{D}^{\alpha}_{\mathbf{r}} \cdot  \!\!\int  \!\rd\mu _{\alpha '} (\mathbf{r}' )
    \int_{-\infty }^{t}\!\!\!\!\!\!\ee^{\varepsilon (t'-t)}
    D^{\alpha\alpha'} (\mathbf{r},\mathbf{r}';t,t' )\cdot \mathrm{D}^{\alpha'}_{\mathbf{r}'}
    \beta \nu   (\mathbf{r}';t')\,\rd t' ,
 \end{equation}
 where
 \begin{equation} \label{02}
  D^{\alpha\alpha'} \!\!(\mathbf{r},\mathbf{r}';t,t')=\big\langle \hat{\mathbf{v}}^{\alpha} (\mathbf{r})T(t,t')\hat{\mathbf{v}}^{\alpha'} (\mathbf{r}')\big\rangle _{\alpha ,rel}^{t}
 \end{equation}
 is the generalized coefficient  diffusion of the particles within the Gibbs statistics.
 Averaging in Eq.~\eqref{02}
 is performed with the power-law Gibbs distribution,
 \begin{equation} \label{03}
  \rho _{rel} (t)=\frac{1}{Z_{G} (t)} \exp \left[- \beta \left(H- \int  \!\rd\mu _{\alpha } (\mathbf{r})\nu   (\mathbf{r};t)\hat{n} (\mathbf{r})\right)\right] ,
 \end{equation}
 where
 \begin{equation} \label{04}
  Z_{G} (t)=\hat{I}^{\alpha } (1,\ldots,N)\hat{T}(1,\ldots,N)
  \exp \left[ - \beta \left(H- \int  \!\rd\mu _{\alpha } (\mathbf{r})\nu  (\mathbf{r};t)\hat{n} (\mathbf{r})\right)\right]
 \end{equation}
 is  the partition function of the relevant distribution function,
 $H$ is a Hamiltonian of the system.
 Parameter $\nu  (\mathbf{r};t)$
 is the chemical potential of the particles,
 which is determined from the self-consistency condition,
 \begin{equation} \label{05}
  \left\langle \hat{n} (\mathbf{r})\right\rangle _{\alpha }^{t} =\left\langle \hat{n} (\mathbf{r})\right\rangle _{\alpha ,rel}^{t} .
 \end{equation}
 $\beta ={1}/{k_{{\rm B}} T} $  ($k_{{\rm B}} $ is the Boltzmann constant),
 $T$ is the equilibrium value of temperature,
 $\hat{\mathbf{v}}^{\alpha} (\mathbf{r})=\sum _{j=1}^{N} \mathbf{v}^{\alpha}_{j} \delta (\mathbf{r}-\mathbf{r}_{j} )$
 is the microscopic flux density of the particles.

 In the Markov approximation,
 the generalized coefficient of  diffusion in time and space has the form
 $D^{\alpha\alpha'} (\mathbf{r},\mathbf{r}';t,t')\approx D \,\delta (t-t')\delta (\mathbf{r}-\mathbf{r}')\delta_{\alpha\alpha'}$.
 And by excluding the parameter $\nu  (\mathbf{r}';t')$ via the self-consistency condition,
 we obtain the diffusion equation with fractional derivatives from Eq.~\eqref{01}
 \begin{equation} \label{GrindEQ__26_}
  \frac{\partial }{\partial t} \left\langle \hat{n} (\mathbf{r})\right\rangle _{\alpha }^{t} = D \,\mathrm{D}^{2\alpha}_{r} \nu (\mathbf{r};t').
 \end{equation}

 The generalized diffusion equation takes into account spatial nonlocality of the system and memory effects in the generalized coefficient of diffusion
 $D^{\alpha\alpha'} (\mathbf{r},\mathbf{r}';t,t')$ within the Gibbs statistics.
 To show the multifractal time in the generalized diffusion equation,
 we use the following approach for the generalized coefficient of particle diffusion
 \begin{equation}
   D^{\alpha\alpha'}(\mathbf r, \mathbf r';t,t')=W(t,t')\overline{D}^{\alpha\alpha'}(\mathbf r, \mathbf r'),
  \end{equation}
 where $W(t,t')$ can be defined as the time memory function.
 In view of this,
 Eq.~\eqref {01} can be represented as
 \begin{equation}\label{eq:2.191}
   \frac{\partial}{\partial t}\left\langle\hat n(\mathbf r)\right\rangle^{t}_{\alpha}
   =\int_{-\infty}^{t}\!\!\ee^{\varepsilon(t'-t)}W(t,t')\Psi(\mathbf r;t')\,\rd t',
 \end{equation}
 where
 \begin{equation}\label{eq:2.192}
   \Psi(\mathbf r;t')
   =\int\! \rd\mu_{\alpha'}(\mathbf r')\,D^{\alpha}_{\mathbf{r}}\cdot \overline{D}^{\alpha\alpha'}(\mathbf r, \mathbf r')\cdot
  D^{\alpha'}_{\mathbf{r}'}\beta\nu(\mathbf r';t').
 \end{equation}

 Further we apply the Fourier transform to Eq.~\eqref{eq:2.191},
 and as a result we get in frequency representation
 \begin{equation}\label{eq:2.193}
  \ri\,\omega\, n(\mathbf r;\omega) =W(\omega)\Psi(\mathbf r;\omega).
 \end{equation}

 We can represent frequency dependence of the memory function  in the following form
 \begin{equation}\label{eq:2.194}
   W(\omega)=\frac{(\ri\,\omega)^{1-\xi}}{1+(\ri\,\omega\, \tau)^{\xi}} ,\quad  0<\xi \leqslant 1,
 \end{equation}
 where the introduced relaxation time $\tau$ characterizes of the particle transport processes in system.
 Then Eq.~\eqref{eq:2.193} can be represented as
 \begin{equation}\label{eq:2.195}
  \big(1+(\ri\,\omega \,\tau)^{\xi}\big)\ri\,\omega \,n(\vec r;\omega) =(\ri\,\omega)^{1-\xi}\Psi(\mathbf r;\omega).
 \end{equation}

 Further we use the Fourier transform to fractional derivatives of functions,
 \begin{equation}\label{eq:2.1096}
   L\big(\DDD{0}{t}{1-\xi}f(t);\ri\,\omega\big)=(\ri\,\omega)^{1-\xi} L\big(f(t);\ri\,\omega\big),
 \end{equation}
 where
 \[
  \DDD{0}{t}{1-\xi}f(t)=\frac{1}{\Gamma(\xi)}\frac{\rd}{\rd t}\int^{t}_{0}
  \frac{f(\tau)}{(t-\tau)^{1-\xi}}\,\rd\tau
 \]
 is the Riemann--Liouville fractional derivative.
 By using it,
 the inverse transformation of Eq.~\eqref{eq:2.195} to time representation gives the Cattaneo--Maxwell generalized diffusion equation
 with  taking into account spatial fractality,
  in the expanded form
 \begin{equation}\label{eq:2.1960}
   \DDD{0}{t}{2\xi} n(\mathbf r;t)\tau^{\xi} +\DDD{0}{t}{\xi} n(\mathbf r;t)
   =\int\! \rd\mu_{\alpha'}(\mathbf r')\,\mathrm{D}_{r}^{\alpha}\cdot \overline{D}(\mathbf r, \mathbf r')\cdot
  \mathrm{D}_{r'}^{\alpha'}\beta\nu(\mathbf r';t),
 \end{equation}
 is the new Cattaneo--Maxwell generalized equation within the Gibbs statistics with  time and spatial nonlocality.
 Eq.~\eqref{eq:2.1960} contains significant spatial heterogeneity in $\overline{D}^{\alpha\alpha'}(\mathbf r, \mathbf r')$.
 If we neglect spatial heterogeneity,
 \begin{equation}\label{eq:2.197}
    \overline{D}^{\alpha\alpha'}(\mathbf r, \mathbf r')= \overline{D}\,\delta (\mathbf r - \mathbf r')\,\delta_{\alpha\alpha'},
 \end{equation}
 we get the Cattaneo--Maxwell diffusion equation with of space-time  nonlocality
 and constant coefficients of  diffusion within the Gibbs statistics
 \begin{equation}\label{eq:2.393}
  \DDD{0}{t}{2\xi} n(\mathbf r;t) \tau^{\xi} +\DDD{0}{t}{\xi} n(\mathbf r;t) =\overline{D}\,\mathrm{D}_{r}^{2\alpha}
    \beta \nu(\mathbf r;t).
 \end{equation}

 From the point of view of the analysis of $n(\mathbf r;t)$ behavior in space and time when the values of diffusion coefficient, characteristic relaxation time and indexes of spatiotemporal fractality are changed, it is important to investigate the frequency spectrum of this equation.
 Peculiarities of dispersion relations are to be expected, since there is a characteristic relaxation time taking into account the indexes of spatiotemporal fractality $\alpha$ $\xi$.
 It is important to investigate the phase $ v_{p}(k)$ and group  $ v_{g}(k)$ velocities when the values of wave vector $k=|\mathbf k|$ are changed,
 because for $v_{p}(k)< v_{g}(k)$ we obtain an anomalous diffusion.

 \section{Dispersion relation for the time-space-fractional Cattaneo--Maxwell diffusion equation}
\setcounter{section}{3}
\setcounter{equation}{0}

 Using the self-consistent condition~\eqref{05} and the approved approximations,
 Eq.~\eqref{eq:2.393} can be written as
 \begin{equation}\label{eq:2.40}
  \DDD{0}{t}{2\xi} n(\mathbf r;t) \tau^{\xi} +\DDD{0}{t}{\xi} n(\mathbf r;t) -\overline{D'}\,\mathrm{D}_{r}^{2\alpha} n(\mathbf r;t)=0,
 \end{equation}
 where $\overline{D'}$ is the renormalized diffusion coefficient.
 For simplicity,
 we consider the one-dimensional case and a solution of Eq.~\eqref{eq:2.40} will be sought in the form of the plane wave,
 $n(x;t)\sim \ee^{-\ri\omega t+\ri kx}$,
 then we get the corresponding frequency spectrum,
 \begin{equation}\label{eq1}
  \tau^{\xi}(-\ri\,\omega)^{2\xi}  + (-\ri\,\omega)^{\xi}-\overline{D'}(\ri\, k)^{2\alpha}=0
 \end{equation}
 Equation~\eqref{eq1} is a quadratic equation in $(-\ri\,\omega)^{\xi}$,
 with discriminant
 \begin{equation}\label{eq2}
   \Delta_{\alpha,\xi}=1+4\overline{D'}\tau^\xi(\ri\, k)^{2\alpha}
 \end{equation}
 and the following roots:
 \begin{equation}\label{eq3}
   (-\ri\,\omega)^{\xi}=\frac{-1\pm\sqrt{\Delta_{\alpha,\xi}}}{2\tau^\xi}.
 \end{equation}

 In the next subsections, for different values of parameters $\alpha$ and $\xi$ the real ($\omega_r(k)=\re \omega(k)$) and imaginary ($\omega_i(k)=\im \omega(k)$) parts of complex frequency ($\omega(k)=\omega_r(k)+\ri\,\omega_i(k)$),
 as well as the phase ($v_p(k)$) and group ($v_g(k)$) velocities will be calculated according to the following definitions:
 \begin{equation}\label{defVpVg}
   v_{p}(k)=\frac{\omega_{r}(k)}{k}, \quad
   v_{g}(k)=\frac{\partial \omega_{r}(k)}{\partial k}.
 \end{equation}

\subsection{Limiting case $\alpha=\xi=1$ (absence of spatial and temporal fractality)}

 In this case, we obtain a dispersion equation for the ordinary Cattaneo--Maxwell equation:
 \begin{equation}\label{DispEqNoNo}
  \tau(-\ri\,\omega)^{2}  + (-\ri\,\omega)+\overline{D'}k^{2}=0,
 \end{equation}
 with discriminant
 \begin{equation}\label{DeltaDispEqNoNo}
   \Delta=1-4\overline{D'}\tau k^{2}
         =1-(k/k_0)^2,
 \end{equation}
 where $k_0=1/\sqrt{4\tau\overline{D'}}$.

 The solution of equation~\eqref{DispEqNoNo},
 which has a physical meaning,
 is well known
 \begin{equation}\label{OmegaRNoNo}
   \omega_{r}(k)=
   \left\{
     \begin{array}{ll}
       0, &                       0\leqslant k \leqslant k_0,\\[2mm]
       \frac1{2\tau}\sqrt{(k/k_0)^{2}-1},  & k > k_0,
     \end{array}
   \right.
 \end{equation}
 \begin{equation}\label{OmegaINoNO}
   \omega_{i}(k)=
   \left\{
     \begin{array}{ll}
       -\frac{1}{2\tau}\left({1+\sqrt{1-(k/k_0)^{2}}}\right), &
                      0\leqslant k \leqslant k_0,\\[3mm]
       -\frac{1}{2\tau},  & k > k_0.
     \end{array}
   \right.
 \end{equation}

 Using the definitions~\eqref{defVpVg} for phase ($v_{p}(k)$) and group ($v_{g}(k)$) velocities,
 we obtain
 that
 \begin{equation}\label{VpNoNO}
   v_{p}(k)=
   \left\{
     \begin{array}{ll}
       0, &                       0\leqslant k \leqslant k_0,\\
        \frac1{2\tau k}\sqrt{(k/k_0)^{2}-1},  & k > k_0,
     \end{array}
   \right.
 \end{equation}
 \begin{equation}\label{VgNoNO}
   v_{g}(k)=
   \left\{
     \begin{array}{ll}
       0, &
                      0\leqslant k \leqslant k_0,\\[2mm]
       \frac{1}{2\tau}\frac{k/k_0^2}{\sqrt{(k/k_0)^2-1}},  & k > k_0.
     \end{array}
   \right.
 \end{equation}

 Note that the real~\eqref{OmegaRNoNo} and imaginary~\eqref{OmegaINoNO} parts of complex frequency,
 as well as the phase velocity~\eqref{VpNoNO} are continuous functions.
 Whereas the group velocity~\eqref{VgNoNO} has a discontinuity of the second kind at the point $k_0$.

\subsection{Limiting case $\alpha=1$ (absence of spatial fractality)}

 In this case, the dispersion equation takes the following form:
 \begin{equation}\label{DispEqXiNo}
  \tau^{\xi}(-\ri\,\omega)^{2\xi}  + (-\ri\,\omega)^{\xi}+\overline{D'}k^{2}=0,
 \end{equation}
 which solution is
 \begin{equation}\label{eq:3.5}
  (-\ri\,\omega)^{\xi} =\frac{-1\pm\sqrt{1-(k/k_{0,\xi})^{2}}}{2\tau^{\xi}},
 \end{equation}
 where $k_{0,\xi}=1/\sqrt{4\tau^{\xi}\overline{D'}}$.

 The real ($\omega_{r}(k)$) and imaginary ($\omega_{i}(k)$) parts of complex frequency are:
 \begin{equation}\label{OmegaRXiNo}
   \omega_{r}(k)=
   \left\{
     \begin{array}{ll}
       -\frac{1}{2^{{1}/{\xi}}\tau}\left[{1\pm\sqrt{1-\left(\frac{k}{k_{0,\xi}}\right)^{2}}}\right]^{\frac{1}{\xi}}\sin\frac{\pi}{\xi}, &
                      0\leqslant k \leqslant k_{0,\xi},\\[4mm]
       \mp \frac1\tau\left(\frac{k}{2k_{0,\xi}}\right)^{\frac{1}{\xi}}
                     \sin\left[-\frac\pi\xi+\frac{1}{\xi}\arctan \sqrt{\left(\frac{k}{k_{0,\xi}}\right)^{2}-1}\right],  & k > k_{0,\xi},
     \end{array}
   \right.
 \end{equation}
 \begin{equation}\label{OmegaIXiNo}
   \omega_{i}(k)=
   \left\{
     \begin{array}{ll}
       \frac{1}{2^{1/\xi}\tau}\left[{1\pm\sqrt{1-\left(\frac{k}{k_{0,\xi}}\right)^{2}}}\right]^{\frac{1}{\xi}}\cos\frac{\pi}{\xi}, &
                      0\leqslant k \leqslant k_{0,\xi},\\[4mm]
       \frac1\tau\left(\frac{k}{2k_{0,\xi}}\right)^{\frac{1}{\xi}}
                     \cos\left[-\frac\pi\xi+\frac{1}{\xi}\arctan \sqrt{\left(\frac{k}{k_{0,\xi}}\right)^{2}-1}\right],  & k > k_{0,\xi}.
     \end{array}
   \right.
 \end{equation}

 According to the definitions~\eqref{defVpVg} for phase ($v_{p}(k)$) and group ($v_{g}(k)$) velocities,
 we obtain
 that
 \begin{equation}\label{VpXiNO}
   v_{p}(k)=
   \left\{
     \begin{array}{ll}
       -\frac{1}{2^{1/\xi}\tau k}\left[{1\pm\sqrt{1-\left(\frac{k}{k_{0,\xi}}\right)^{2}}}\right]^{\frac{1}{\xi}}\sin\frac{\pi}{\xi}, &
                      0\leqslant k \leqslant k_{0,\xi},\\[4mm]
       \mp \frac1{\tau k}\left(\frac{k}{2k_{0,\xi}}\right)^{\frac{1}{\xi}}
                     \sin\left[-\frac\pi\xi+\frac{1}{\xi}\arctan \sqrt{\left(\frac{k}{k_{0,\xi}}\right)^{2}-1}\right],  & k > k_{0,\xi},
     \end{array}
   \right.
 \end{equation}
 \begin{equation}\label{VgXiNO}
   v_{g}(k)\!=\!
   \left\{\!\!
     \begin{array}{ll}
       \pm \frac{1}{2^{1/\xi}\tau \xi} \frac{k/k_{0,\xi}^2}{\sqrt{1-(k/k_{0,\xi})^{2}}}
   \left[{1\pm\sqrt{1-\left(\frac{k}{k_{0,\xi}}\right)^{2}}}\right]^{\frac{1-\xi}{\xi}}\!\!\!\sin\frac{\pi}{\xi}, &
              \!\!\!        0\leqslant k \leqslant k_{0,\xi},\\[4mm]
        \mp \frac{1}{\xi\tau k} \left({\frac{k}{2k_{0,\xi}}}\right)^{\frac{1}{\xi}} \!\!
   \Bigg\{\sin\left[-\frac\pi\xi+\frac{1}{\xi}\arctan \sqrt{\left(\frac{k}{k_{0,\xi}}\right)^{2}-1}\right]  \\[4mm]
           +\frac{1}{\sqrt{(k/k_{0,\xi})^{2}-1}}\cos\left[-\frac\pi\xi+\frac{1}{\xi}\arctan \sqrt{\left(\frac{k}{k_{0,\xi}}\right)^{2}-1}\right]\Bigg\},  &
        \!\!\!   k > k_{0,\xi}.
     \end{array}
   \right.
 \end{equation}

 Note that
 \[
   \lim_{k\to k_{0,\xi}-0}\omega_r(k) =-\frac{1}{2^{1/\xi}\tau}\sin\frac{\pi}{\xi}, \quad
   \lim_{k\to k_{0,\xi}+0}\omega_r(k)=\mp\frac{1}{2^{1/\xi}\tau}\sin\frac{\pi}{\xi},
 \]
 \[
  \lim_{k\to k_{0,\xi}-0}\omega_i(k) =\lim_{k\to k_0+0}\omega_i(k) =\frac{1}{2^{1/\xi}\tau}\cos\frac{\pi}{\xi},
 \]
 \[
   \lim_{k\to k_{0,\xi}-0}v_p(k)=-\frac{1}{2^{1/\xi}\tau k}\sin\frac{\pi}{\xi}, \quad
   \lim_{k\to k_{0,\xi}+0}v_p(k)=\mp\frac{1}{2^{1/\xi}\tau k}\sin\frac{\pi}{\xi}.
 \]
 That is, one branch of the real part of complex frequency is continuous and the other has a discontinuity of the first kind at the point $k=k_0$ (and accordingly the phase velocity);
 whereas the imaginary part of frequency is everywhere continuous function.

 Note that in the limit $\xi\to1$ expressions~\eqref{OmegaRXiNo}--\eqref{VgXiNO} must turn into expressions~\eqref{OmegaRNoNo}--\eqref{VgNoNO}, respectively.
 This allows one to define the signs in expressions~\eqref{OmegaRNoNo}--\eqref{VgNoNO}.
 Thus, for the imaginary part of complex frequency~\eqref{OmegaIXiNo},
 as well as for the real part of complex frequency~\eqref{OmegaRXiNo} for $k > k_{0,\xi}$,
 and accordingly for the phase and group velocities, one needs to take the upper sign.
 Whereas for the real part of complex frequency~\eqref{OmegaRXiNo} for $k \leqslant k_{0,\xi}$,
 and accordingly for the phase and group velocities, there are both signs,
 because this real part, and accordingly the phase and group velocities,
 becomes zero ($k \leqslant k_{0,\xi}$) for both signs.
 As a result, in the case of $\alpha=1$ and $\xi\neq1$ in the real part of complex frequency
 (and accordingly in the phase and group velocities) a bifurcation appears at the point $k=k_{0,\xi}$,
 to the left of which there are two branches.

\subsection{Limiting case $\xi=1$ (absence of temporal fractality)}

 In this case, the dispersion equation takes the following form:
 \begin{equation}\label{DispEqAlfaNo}
  \tau(-\ri\,\omega)^{2}  + (-\ri\,\omega)-\overline{D'}(\ri\, k)^{2\alpha}=0,
 \end{equation}
 the discriminant of which is:
 \begin{equation}\label{eq5}
   \Delta_\alpha=1+4\overline{D'}\tau(\ri\, k)^{2\alpha},
 \end{equation}
 the roots are:
 \begin{equation}\label{eq6}
   -\ri\,\omega=\frac{-1\pm\sqrt{\Delta_\alpha}}{2\tau},
 \end{equation}
 where the following notations are introduced:
 \[
  \Delta_\alpha=|\Delta_\alpha|\,\ee^{\ri\psi},\quad
  |\Delta_\alpha|=\sqrt{1+2\left(\frac{k}{k_{0,\alpha}}\right)^{2\alpha}\cos(\alpha\pi)+\left(\frac{k}{k_{0,\alpha}}\right)^{4\alpha}},
 \]
 \[
  \psi=\arctan\frac{\left(\frac{k}{k_{0,\alpha}}\right)^{2\alpha}\sin(\alpha\pi)}{1+\left(\frac{k}{k_{0,\alpha}}\right)^{2\alpha}\cos(\alpha\pi)},\quad
  k_{0,\alpha}=\frac{1}{(4\overline{D'}\tau)^{\frac1{2\alpha}}}.
 \]

 If $0\leqslant\alpha\leqslant\frac{1}{2}$,
 then the real and imaginary parts of complex frequency are as follows:
 \begin{equation}\label{OmegaRAlfaNo1}
   \textstyle \omega_r(k)=\pm\frac{1}{2\tau}|\Delta_\alpha|^{\frac12}\sin\frac{\psi}{2},
 \end{equation}
 \begin{equation}\label{OmegaIAlfaNo1}
   \textstyle \omega_i(k)=   \frac{1}{2\tau}\left(-1\pm|\Delta_\alpha|^{\frac12}\cos\frac{\psi}{2}\right).
 \end{equation}
 Then, according to the definitions~\eqref{defVpVg}, the phase and group velocities are as follows:
 \begin{equation}\label{VpAlfaNo1}
  \textstyle v_p(k)=\pm\frac{1}{2\tau k}|\Delta_\alpha|^{\frac12}\sin\frac{\psi}{2},
 \end{equation}
 \begin{equation}\label{VgAlfaNo1}
  \textstyle v_g(k)=\pm\frac{\alpha}{2\tau k|\Delta_\alpha|^{\frac32}}
         \left(\frac{k}{k_{0,\alpha}}\right)^{2\alpha}\left[
          \left(\frac{k}{k_{0,\alpha}}\right)^{2\alpha}\!\!\sin\frac{\psi}{2}
           +\sin\left(\alpha\pi+\frac{\psi}{2}\right)
         \right].
 \end{equation}

 If $\frac{1}{2}<\alpha\leqslant1$,
 then the real and imaginary parts of complex frequency are as follows:
 \begin{equation}\label{OmegaRAlfaNo}
   \omega_r(k)=  \left\{
    \begin{array}{ll}
      \pm\frac{1}{2\tau}|\Delta_\alpha|^{\frac12}\sin\frac{\psi}{2}, & 0\leqslant k\leqslant \frac{k_{0,\alpha}}{(-\cos(\alpha\pi))^{\frac{1}{2\alpha}}}, \\[3mm]
      \pm\frac{1}{2\tau}|\Delta_\alpha|^{\frac12}\cos\frac{\psi}{2}, & k>\frac{k_{0,\alpha}}{(-\cos(\alpha\pi))^{\frac{1}{2\alpha}}},
    \end{array}
  \right.
 \end{equation}
 \begin{equation}\label{OmegaIAlfaNo}
   \omega_i(k)=     \left\{
    \begin{array}{ll}
     \frac{1}{2\tau}\left(-1\pm|\Delta_\alpha|^{\frac12}\cos\frac{\psi}{2}\right), & 0\leqslant k\leqslant \frac{k_{0,\alpha}}{(-\cos(\alpha\pi))^{\frac{1}{2\alpha}}}, \\[3mm]
     \frac{1}{2\tau}\left(-1\mp|\Delta_\alpha|^{\frac12}\sin\frac{\psi}{2}\right), & k>\frac{k_{0,\alpha}}{(-\cos(\alpha\pi))^{\frac{1}{2\alpha}}}.
    \end{array}
  \right.
 \end{equation}
Then, according to the definitions~\eqref{defVpVg}, the phase and group velocities are as follows:
 \begin{equation}\label{VpAlfaNo}
  v_p(k)=
  \left\{
    \begin{array}{ll}
      \pm\frac{1}{2\tau k}|\Delta_\alpha|^{\frac12}\sin\frac{\psi}{2}, & 0\leqslant k\leqslant \frac{k_{0,\alpha}}{(-\cos(\alpha\pi))^{\frac{1}{2\alpha}}}, \\[2mm]
      \pm\frac{1}{2\tau k}|\Delta_\alpha|^{\frac12}\cos\frac{\psi}{2}, & \frac{k_{0,\alpha}}{(-\cos(\alpha\pi))^{\frac{1}{2\alpha}}}< k<\infty,
    \end{array}
  \right.
 \end{equation}
 \begin{equation}\label{VgAlfaNo}
  v_g(k)=
  \left\{
    \begin{array}{l}
      \pm\frac{\alpha}{2\tau k|\Delta_\alpha|^{\frac32}}
         \left(\frac{k}{k_{0,\alpha}}\right)^{2\alpha}\left[
          \left(\frac{k}{k_{0,\alpha}}\right)^{2\alpha}\!\sin\frac{\psi}{2}
           +\sin\left(\alpha\pi+\frac{\psi}{2}\right)
         \right],  \\
        \qquad  0\leqslant k\leqslant \frac{k_{0,\alpha}}{(-\cos(\alpha\pi))^{\frac{1}{2\alpha}}}, \\[5mm]
      \pm\frac{\alpha}{2\tau k|\Delta_\alpha|^{\frac32}}
         \left(\frac{k}{k_{0,\alpha}}\right)^{2\alpha}\left[
          \left(\frac{k}{k_{0,\alpha}}\right)^{2\alpha}\!\cos\frac{\psi}{2}
           +\cos\left(\alpha\pi+\frac{\psi}{2}\right)
         \right],  \\
       \qquad  \frac{k_{0,\alpha}}{(-\cos(\alpha\pi))^{\frac{1}{2\alpha}}}< k<\infty.
    \end{array}
  \right.
 \end{equation}

 Note that
 \[
  \lim_{k\to \frac{k_{0,\alpha}}{(-\cos(\alpha\pi))^{\frac{1}{2\alpha}}}\mp0}\psi
  =\lim_{k\to \frac{k_{0,\alpha}}{(-\cos(\alpha\pi))^{\frac{1}{2\alpha}}}\mp0}
    \arctan\frac{\left(\frac{k}{k_{0,\alpha}}\right)^{2\alpha}\sin(\alpha\pi)}{1+\left(\frac{k}{k_{0,\alpha}}\right)^{2\alpha}\cos(\alpha\pi)}
  =\pm\frac{\pi}{2}
 \]
 and
 \begin{align*}
  \lim_{k\to \frac{k_{0,\alpha}}{(-\cos(\alpha\pi))^{\frac{1}{2\alpha}}}}|\Delta_\alpha|
  &=\lim_{k\to \tfrac{k_{0,\alpha}}{(-\cos(\alpha\pi))^{\frac{1}{2\alpha}}}}
    \sqrt{1+2\left(\tfrac{k}{k_{0,\alpha}}\right)^{2\alpha}\cos(\alpha\pi)+\left(\tfrac{k}{k_{0,\alpha}}\right)^{4\alpha}}
    \\
  &=|\tan(\alpha\pi)|.
 \end{align*}
 So the real~\eqref{OmegaRAlfaNo} and imaginary~\eqref{OmegaIAlfaNo} parts of complex frequency, as well as
 the phase~\eqref{VpAlfaNo} and group~\eqref{VgAlfaNo} velocities are continuous functions.

 Note that in the limit $\alpha\to1$
 expressions~\eqref{OmegaRAlfaNo}--\eqref{VgAlfaNo} must turn into expressions~\eqref{OmegaRNoNo}--\eqref{VgNoNO}, respectively.
 Since
 \[
  \lim_{\alpha\to1}\psi
  =\lim_{\alpha\to1}
    \arctan\frac{\left(\frac{k}{k_{0,\alpha}}\right)^{2\alpha}\sin(\alpha\pi)}{1+\left(\frac{k}{k_{0,\alpha}}\right)^{2\alpha}\cos(\alpha\pi)}=0
 \]
 and
 \[
  \lim_{\alpha\to1}|\Delta_\alpha|
  =\lim_{\alpha\to1}
    \sqrt{1+2\left(\tfrac{k}{k_{0,\alpha}}\right)^{2\alpha}\cos(\alpha\pi)+\left(\tfrac{k}{k_{0,\alpha}}\right)^{4\alpha}}
  =\left|1-\left(\tfrac{k}{k_{0}}\right)^{2}\right|,
 \]
 then
 \begin{equation}\label{OmegaRNoNo2}
   \omega_{r}(k)=
   \left\{
     \begin{array}{ll}
       0, &                       0\leqslant k \leqslant k_0,\\[2mm]
       \pm\frac1{2\tau}\sqrt{\left({k}/{k_{0}}\right)^{2}-1},  & k > k_0,
     \end{array}
   \right.
 \end{equation}
 \begin{equation}\label{OmegaINoNO2}
   \omega_{i}(k)=
   \left\{
     \begin{array}{ll}
       -\frac{1}{2\tau}\left[{1\pm\sqrt{1-\left({k}/{k_{0}}\right)^{2}}}\right], &
                      0\leqslant k \leqslant k_0,\\[3mm]
       -\frac{1}{2\tau},  & k > k_0,
     \end{array}
   \right.
 \end{equation}
 \begin{equation}\label{VpNoNO2}
   v_{p}(k)=
   \left\{
     \begin{array}{ll}
       0, &                       0\leqslant k \leqslant k_0,\\
       \pm \frac1{2\tau k}\sqrt{\left({k}/{k_{0}}\right)^{2}-1},  & k > k_0,
     \end{array}
   \right.
 \end{equation}
 \begin{equation}\label{VgNoNO2}
   v_{g}(k)=
   \left\{
     \begin{array}{ll}
       0, &
                      0\leqslant k \leqslant k_0,\\[2mm]
       \pm\frac{1}{2\tau}\frac{k/k_0^2}{\sqrt{(k/k_0)^2-1}},  & k > k_0.
     \end{array}
   \right.
 \end{equation}
 It follows, from the comparison of expressions~\eqref{OmegaRNoNo2}--\eqref{VgNoNO2} with the corresponding expressions~\eqref{OmegaRNoNo}--\eqref{VgNoNO}, that one needs to choose the ``+'' sign in expressions~\eqref{OmegaRAlfaNo}--\eqref{VgAlfaNo}.

 The signs in expressions~\eqref{OmegaRAlfaNo1}--\eqref{VgAlfaNo1} must be chosen so, that in the limit $\alpha\to1/2$ these expressions turn into expressions~\eqref{OmegaRAlfaNo}--\eqref{VgAlfaNo} for $\alpha=1/2$.
 To do this, one needs also to choose the ``+'' sign in expressions~\eqref{OmegaRAlfaNo1}--\eqref{VgAlfaNo1}.

\begin{figure}[htbp]
  \centering
  \includegraphics[width=0.5\textwidth]{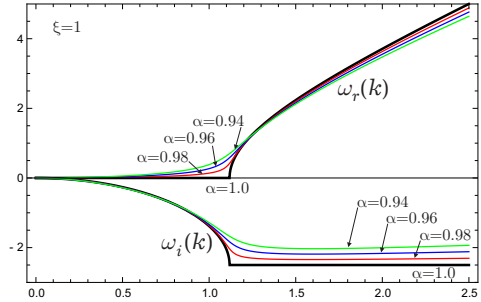}\\
  \caption{Frequency spectrum for $\xi=1$ and $\alpha=1.0$, $0.98$, $0.96$, $0.94$ ($\tau=0.2$ and $\overline{D'}=1$).\label{OmegaXi1}}
\end{figure}

 Fig.\,\ref{OmegaXi1} shows the  dependencies of the real ($\omega_r(k)$)~\eqref{OmegaRAlfaNo} and imaginary ($\omega_i(k)$)~\eqref{OmegaIAlfaNo} parts of complex frequency for $\xi=1$ and $\alpha=1.0$, $0.98$, $0.96$, $0.94$.
 Black bold lines represent dependencies~\eqref{OmegaRNoNo} and \eqref{OmegaINoNO}, that is, when $\alpha=1$.
 In the case of  $\alpha\neq1$ these dependencies are smoothed, and
 when approaching $\alpha\to1$ they converge uniformly to expressions~\eqref{OmegaRNoNo} and \eqref{OmegaINoNO} for the  ordinary Cattaneo--Maxwell equation.

\begin{figure}[htbp]
  \centering
  \includegraphics[width=0.5\textwidth]{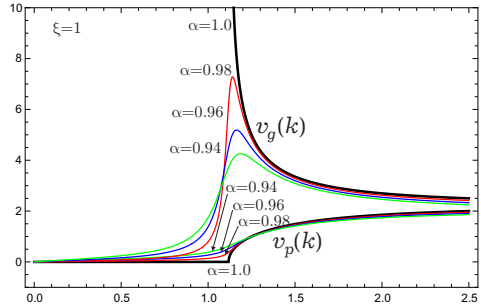}\\
  \caption{Phase ($v_p(k)$) and group ($v_g(k)$) velocities for  $\xi=1$ and $\alpha=1.0$, $0.98$, $0.96$, $0.94$ ($\tau=0.2$ and $\overline{D'}=1$).\label{VXi1}}
\end{figure}

 Fig.\,\ref{VXi1} shows the dependencies of the phase ($v_p(k)$)~\eqref{VpAlfaNo} and group ($v_g(k)$)~\eqref{VgAlfaNo} velocities for $\xi=1$ and $\alpha=1.0$, $0.98$, $0.96$, $0.94$.
 Black bold lines represent dependencies~\eqref{VpNoNO} and \eqref{VgNoNO},
 that is, when $\alpha=1$.
 In this case the group velocity ($v_g(k)$) has a discontinuity.
 In the case of $\alpha\neq1$ these dependencies are smoothed,
 instead of a discontinuity the maximum is observed, the value of which decreases with decreasing $\alpha$.
 When approaching $\alpha\to1$ the phase and group velocities converge uniformly to expressions~\eqref{VpNoNO} and \eqref{VgNoNO} for the  ordinary Cattaneo--Maxwell equation.

 \subsection{General case $0<\alpha\leqslant1$, $0<\xi\leqslant1$}

\begin{figure}[htbp]
  \centering
  \includegraphics[width=\textwidth]{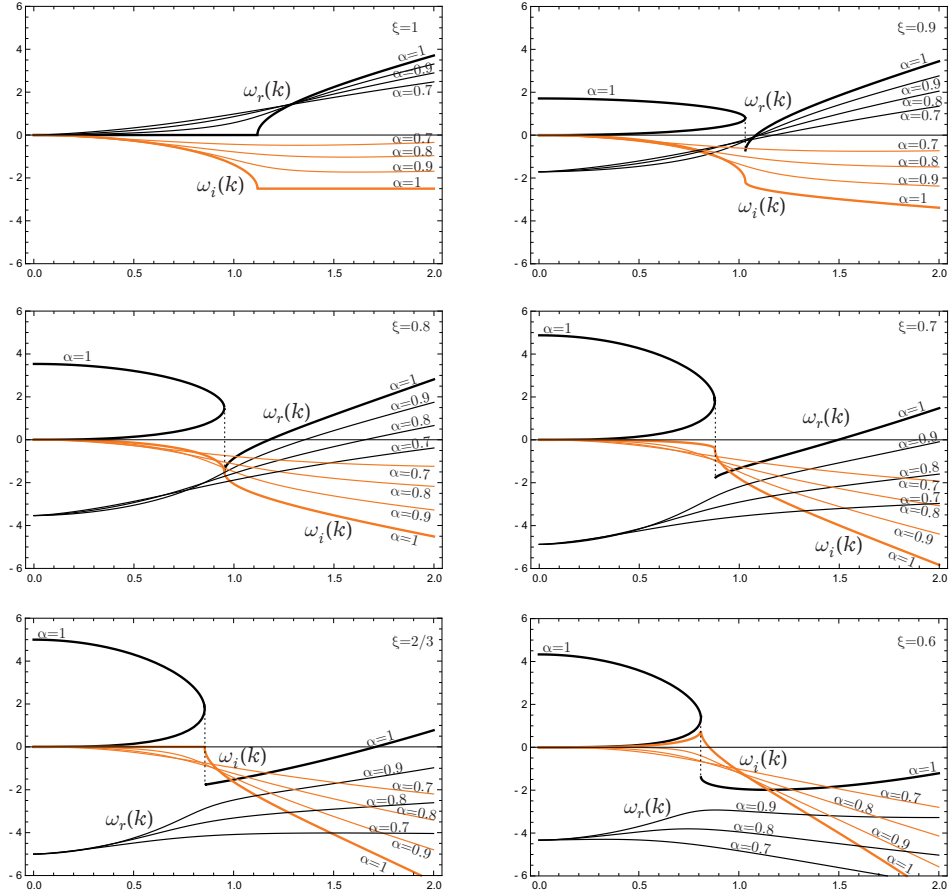}\\
  \caption{Frequency spectrum for $\tau=0.2$ and $\overline{D'}=1$.\label{OM}}
\end{figure}

\begin{figure}[htbp]
  \centering
  \includegraphics[width=\textwidth]{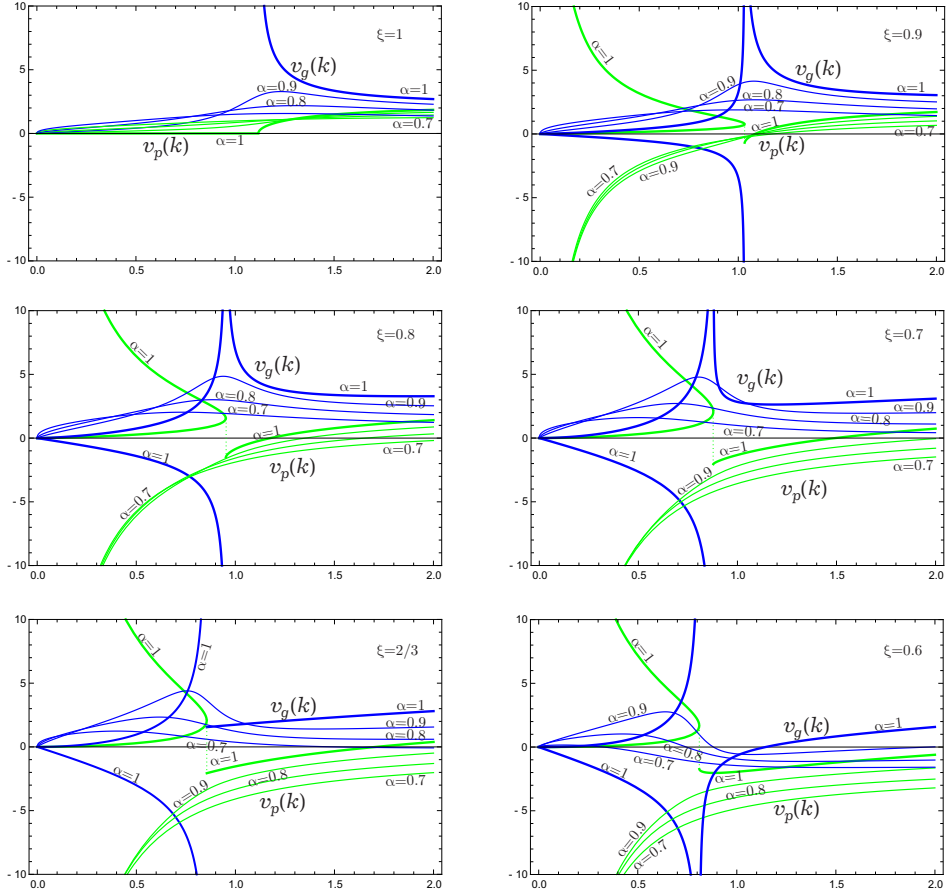}\\
  \caption{Phase ($v_p(k)$) and group ($v_g(k)$) velocities for $\tau=0.2$ and $\overline{D'}=1$.\label{VGP}}
\end{figure}

 Unfortunately, in this case analytical expressions cannot be found,
 therefore, the analysis was performed numerically (see Figs.\,\ref{OM},\,\ref{VGP}).

 Fig.\,\ref{OM} shows the real ($\omega_r(k)$) and imaginary ($\omega_i(k)$) parts of complex frequency,
 and Fig.\,\ref{VGP} shows the phase ($v_p(k)$) and group ($v_g(k)$) velocities as functions of the wave number $k$ for different values of fractality parameters $\alpha$ and $\xi$.

 If $\alpha=1$ and $\xi=1$, then we observe the known dependences~\eqref{OmegaRNoNo}-- \eqref{VgNoNO} for the
 ordinary Cattaneo--Maxwell equation.
 Decreasing the parameter $\alpha$ leads to smoothing of these dependencies.

 If $\alpha=1$ and $\xi\neq1$, then in the real part of complex frequency a bifurcation and discontinuity of the first kind are observed at the point $k \leqslant k_{0,\xi}$, to the left of which two branches exist.
 Accordingly, the phase velocity also has a bifurcation and discontinuity of the first kind at the point $k = k_{0,\xi}$, and the group velocity has a singularity at this point and shows the $\lambda$-like behavior.
 The imaginary part of complex frequency is continuous and has an inflection at this point.

 If $\xi\neq1$ and $\alpha\neq1$, then the bifurcation disappears, the real part of complex frequency is continuous.
 Accordingly, the phase velocity also becomes continuous, and the group velocity doesn't have any singularities and its lower branch (which was present for $\alpha=1$) disappears.
 As parameter $\alpha$ decreases, the dependencies of the real and imaginary parts of complex frequency,  phase and group velocities are smoothed.

 For $\xi=2/3$ and $\alpha=1$ the imaginary part of complex frequency equals zero in the domain $k\leqslant k_{0,\xi}$, and the real part of complex frequency becomes convex downward in the domain $k> k_{0,\xi}$, this causes a significant change in the group velocity behavior: the $\lambda$-like behavior is lost.
 The further decreasing of $\xi$ for $\alpha=1$ leads to an increase in this downward convexity of the real part of complex frequency in the domain   $k> k_{0,\xi}$ and a significant change in the group velocity:
 it becomes $\lambda$-like again, but inverted,
 moreover,
 a domain appears where the imaginary part of complex  frequency becomes positive.
 For $\alpha\neq1$ the bifurcation of the real part of complex frequency disappears,
 the dependencies of the real and imaginary parts of complex frequency are smoothed.

\section{Conclusions}

 When describing non-Markov diffusion processes with spatiotemporal nonlocality of the diffusion coefficient, problems arise in calculating the time correlation function ``velocity--velocity'' $\langle\mathbf{v}(\mathbf{r};t)\mathbf{v}(\mathbf{r}';t')\rangle$.
 In general, for such a calculation, one can use the method of Mori projection operators and express it through higher memory functions.
 But for systems, which have certain characteristic relaxation times or spatial characteristics, fractality, these characteristics must be taken into account.
 We proceeded from the non-Markov diffusion equation taking into account the spatial fractality and modeled the generalized coefficient of particle diffusion as follows: $D^{\alpha\alpha'}(\mathbf r, \mathbf r';t,t')=W(t,t')\overline{D}^{\alpha\alpha'}(\mathbf r, \mathbf r')$.

 Using fractional calculus and the corresponding approximation for time memory function $W(t,t')$ with introduction of the characteristic relaxation time, we have obtained the generalized Cattaneo--Maxwell--type diffusion equation in fractional time and space derivatives.
 In the case of a constant diffusion coefficient,
 analytical and numerical studies of the frequency spectrum for the Cattaneo--Maxwell diffusion equation in fractional time and space derivatives have been performed.
 Numerical calculations of the phase and group velocities with change of values of characteristic relaxation time, diffusion coefficient and indexes of temporal $\xi$ and spatial $\alpha$ fractality have been carried out.
 Calculations have shown, that
 in the case of $\alpha=1$ and $\xi\neq1$ in the real part of complex frequency
 (and accordingly in the phase and group velocities) a bifurcation appears at the point $k=k_{0,\xi}$,
 to the left of which there are two branches.

 In the case when $\alpha=1$ and $\xi=1$  there is a discontinuity in the gruop velocity ($v_g(k)$), which corresponds to the ordinary Cattaneo--Maxwell equation.
 In the case of $\alpha\neq1$ and $\xi=1$ these dependencies are smoothed, insted of a discontinuity the maximum is observed, the value of which decreases with decreasing $\alpha$.
 When approaching $\alpha\to1$ the phase and group velocities converge uniformly to expressions~\eqref{VpNoNO} and \eqref{VgNoNO} for the  ordinary Cattaneo--Maxwell equation.
 In the general case, when $\xi\neq1$ and $\alpha\neq1$, the bifurcation disappears, the real part of complex frequency is continuous.
 Accordingly, the phase velocity also becomes continuous, and the group velocity doesn't have any singularities and its lower branch (which was present for $\alpha=1$) disappears.
 As parameter $\alpha$ decreases, the dependencies of the real and imaginary parts of complex frequency,  phase and group velocities are smoothed.

 The domain is special for $\xi=2/3$ and $\alpha=1$, when the imaginary part of complex frequency equals zero in the domain $k\leqslant k_{0,\xi}$,
 and the real part of complex frequency becomes convex downward in the domain $k> k_{0,\xi}$.
 This causes a significant change in the group velocity behavior: the $\lambda$-like behavior is lost.
 The further decreasing of $\xi$ for $\alpha=1$ leads to an increase in this downward convexity of the real part of complex frequency in the domain   $k> k_{0,\xi}$ and a significant change in the group velocity:
 it becomes $\lambda$-like again, but inverted.
 Moreover,
 a domain appears where the imaginary part of complex  frequency becomes positive.


\begin{thebibliography}{100}

\bibitem{Oldham2006}
K.~B. Oldham and J.~Spanier.
\newblock \emph{The Fractional Calculus: Theory and Applications of
  Differentiation and Integration to Arbitrary Order}.
\newblock Dover Books on Mathematics. Dover Publications (2006).

\bibitem{Samko1993}
S.~G. Samko, A.~A. Kilbas, and O.~I. Marichev.
\newblock \emph{Fractional Integrals and Derivatives: Theory and Applications}.
\newblock Gordon and Breach Science Publishers, 1 edition (1993).

\bibitem{Podlubny1998}
I.~Podlubny and V.~T.~E. Kenneth.
\newblock \emph{Fractional Differential Equations: An Introduction to
  Fractional Derivatives, Fractional Differential Equations, to Methods of
  Their Solution and Some of Their Applications}.
\newblock Mathematics in Science and Engineering 198. Academic Press, 1st
  edition (1998).

\bibitem{Mandelbrot1982}
B.~B. Mandelbrot.
\newblock \emph{The fractal geometry of nature}.
\newblock W. H. Freeman and Company (1982).

\bibitem{Uchaikin2008500}
V.~V. Uchaikin.
\newblock \emph{Fractional Derivatives Method}.
\newblock Artishock-Press, Uljanovsk (2008).

\bibitem{Sahimi1998213}
M.~Sahimi.
\newblock Non-linear and non-local transport processes in heterogeneous media:
  from long-range correlated percolation to fracture and materials breakdown.
\newblock \emph{Physics Reports} \textbf{306}, no. 4--6, (1998), 213--395.
\newblock \doi{https://doi.org/10.1016/S0370-1573(98)00024-6}.

\bibitem{Korosak20071}
D.~Koro\u{s}ak, B.~Cvikl, J.~Kramer, R.~Jecl, and A.~Prapotnik.
\newblock Fractional calculus applied to the analysis of spectral electrical
  conductivity of clay–water system.
\newblock \emph{Journal of Contaminant Hydrology} \textbf{92}, no. 1–-2,
  (2007), 1--9.
\newblock \doi{http://dx.doi.org/10.1016/j.jconhyd.2006.11.005}.

\bibitem{Metzler20001}
R.~Metzler and J.~Klafter.
\newblock The random walk's guide to anomalous diffusion: a fractional dynamics
  approach.
\newblock \emph{Physics Reports} \textbf{339}, no.~1, (2000), 1--77.
\newblock \doi{http://dx.doi.org/10.1016/S0370-1573(00)00070-3}.

\bibitem{Hilfer2000}
R.~Hilfer.
\newblock \emph{Fractional Time Evolution}, chapter~II.
\newblock World Scientific, Singapore, New Jersey, London, Hong Kong, 87--130
  (2000).

\bibitem{Bisquert20002287}
J.~Bisquert, G.~Garcia-Belmonte, F.~Fabregat-Santiago, N.~S. Ferriols,
  P.~Bogdanoff, and E.~C. Pereira.
\newblock {Doubling Exponent Models for the Analysis of Porous Film Electrodes
  by Impedance. Relaxation of TiO$_2$ Nanoporous in Aqueous Solution}.
\newblock \emph{The Journal of Physical Chemistry B} \textbf{104}, no.~10,
  (2000), 2287--2298.
\newblock \doi{https://doi.org/10.1021/jp993148h}.

\bibitem{Bisquert2001112}
J.~Bisquert and A.~Compte.
\newblock Theory of the electrochemical impedance of anomalous diffusion.
\newblock \emph{Journal of Electroanalytical Chemistry} \textbf{499}, no.~1,
  (2001), 112--120.
\newblock \doi{http://dx.doi.org/10.1016/S0022-0728(00)00497-6}.

\bibitem{Kosztolowicz2009055004}
T.~Koszto\l{}owicz and K.~D. Lewandowska.
\newblock Hyperbolic subdiffusive impedance.
\newblock \emph{Journal of Physics A: Mathematical and Theoretical}
  \textbf{42}, no.~5, (2009), 055004.
\newblock \doi{https://doi.org/10.1088/1751-8113/42/5/055004}.

\bibitem{Pyanylo201484}
Y.~D. Pyanylo, M.~G. Prytula, N.~M. Prytula, and N.~B. Lopuh.
\newblock Models of mass transfer in gas transmission systems.
\newblock \emph{Mathematical Modeling and Computing} \textbf{1}, no.~1, (2014),
  84--96.

\bibitem{Zhokh20187176}
A.~Zhokh, A.~Trypolskyi, and P.~Strizhak.
\newblock Relationship between the anomalous diffusion and the fractal
  dimension of the environment.
\newblock \emph{Chemical Physics} \textbf{503}, (2018), 71--76.
\newblock \doi{https://doi.org/10.1016/j.chemphys.2018.02.015}.

\bibitem{Zhokh201735}
A.~A. Zhokh and P.~E. Strizhak.
\newblock Effect of zeolite ZSM-5 content on the methanol transport in the
  ZSM-5/alumina catalysts for methanol-to-olefin reaction.
\newblock \emph{Chemical Engineering Research and Design} \textbf{127}, (2017),
  35--44.
\newblock \doi{https://doi.org/10.1016/j.cherd.2017.09.010}.

\bibitem{Zhokh2017124704}
A.~Zhokh and P.~Strizhak.
\newblock Non-Fickian diffusion of methanol in mesoporous media: Geometrical
  restrictions or adsorption-induced?
\newblock \emph{The Journal of Chemical Physics} \textbf{146}, no.~12, (2017),
  124704.
\newblock \doi{https://doi.org/10.1063/1.4978944}.

\bibitem{Scher19752455}
H.~Scher and E.~W. Montroll.
\newblock Anomalous transit-time dispersion in amorphous solids.
\newblock \emph{Phys. Rev. B} \textbf{12}, no.~6, (1975), 2455--2477.
\newblock \doi{https://doi.org/10.1103/PhysRevB.12.2455}.

\bibitem{Berkowitz19985858}
B.~Berkowitz and H.~Scher.
\newblock Theory of anomalous chemical transport in random fracture networks.
\newblock \emph{Phys. Rev. E} \textbf{57}, no.~5, (1998), 5858--5869.
\newblock \doi{https://doi.org/10.1103/PhysRevE.57.5858}.

\bibitem{Bouchaud1990127}
J.-P. Bouchaud and A.~Georges.
\newblock Anomalous diffusion in disordered media: Statistical mechanisms,
  models and physical applications.
\newblock \emph{Physics Reports} \textbf{195}, no.~4, (1990), 127--293.
\newblock \doi{http://dx.doi.org/10.1016/0370-1573(90)90099-N}.

\bibitem{Nigmatullin1984739}
R.~R. Nigmatullin.
\newblock {To the Theoretical Explanation of the ``Universal Response''}.
\newblock \emph{Physica Status Solidi (B)} \textbf{123}, no.~2, (1984),
  739--745.
\newblock \doi{https://doi.org/10.1002/pssb.2221230241}.

\bibitem{Nigmatullin1984389}
R.~R. Nigmatullin.
\newblock {On the Theory of Relaxation for Systems with ``Remnant'' Memory}.
\newblock \emph{Physica Status Solidi (B)} \textbf{124}, no.~1, (1984),
  389--393.
\newblock \doi{https://doi.org/10.1002/pssb.2221240142}.

\bibitem{Nigmatullin1986425}
R.~R. Nigmatullin.
\newblock The realization of the generalized transfer equation in a medium with
  fractal geometry.
\newblock \emph{Physica Status Solidi (B)} \textbf{133}, no.~1, (1986),
  425--430.
\newblock \doi{https://doi.org/10.1002/pssb.2221330150}.

\bibitem{Nigmatullin1992242}
R.~R. Nigmatullin.
\newblock Fractional integral and its physical interpretation.
\newblock \emph{Theoretical and Mathematical Physics} \textbf{90}, no.~3,
  (1992), 242--251.
\newblock \doi{https://doi.org/10.1007/BF01036529}.

\bibitem{Nigmatullin199787}
R.~R. Nigmatullin and Y.~E. Ryabov.
\newblock {Cole-Davidson} dielectric relaxation as a self-similar relaxation
  process.
\newblock \emph{Physics of the Solid State} \textbf{39}, no.~1, (1997), 87--90.
\newblock \doi{https://doi.org/10.1134/1.1129804}.

\bibitem{Nigmatullin2009014001}
R.~R. Nigmatullin.
\newblock Dielectric relaxation phenomenon based on the fractional kinetics:
  theory and its experimental confirmation.
\newblock \emph{Physica Scripta} \textbf{2009}, no. T136, (2009), 014001.
\newblock \doi{https://doi.org/10.1088/0031-8949/2009/T136/014001}.

\bibitem{Khamzin20121604}
A.~A. Khamzin, R.~R. Nigmatullin, and I.~I. Popov.
\newblock Microscopic model of a non-Debye dielectric relaxation: {The
  Cole-Cole} law and its generalization.
\newblock \emph{Theoretical and Mathematical Physics} \textbf{173}, no.~2,
  (2012), 1604--1619.
\newblock \doi{https://doi.org/10.1007/s11232-012-0135-1}.

\bibitem{Popov20121}
I.~I. Popov, R.~R. Nigmatullin, E.~Y. Koroleva, and A.~A. Nabereznov.
\newblock The generalized {Jonscher's} relationship for conductivity and its
  confirmation for porous structures.
\newblock \emph{Journal of Non-Crystalline Solids} \textbf{358}, no.~1, (2012),
  1--7.
\newblock \doi{http://dx.doi.org/10.1016/j.jnoncrysol.2011.07.020}.

\bibitem{Grygorchak2015e}
I.~I. Grygorchak, P.~P. Kostrobij, I.~V. Stasjuk, M.~V. Tokarchuk, O.~V.
  Velychko, F.~O. Ivaschyshyn, and B.~M. Markovych.
\newblock \emph{Fizichni procesy ta ih mikroskopichni modeli v periodychnyh
  neorganichno/organichnih klatratah}.
\newblock Rastr-7, Lviv (2015).

\bibitem{Kostrobij2015154}
P.~P. Kostrobij, I.~I. Grygorchak, F.~O. Ivaschyshyn, B.~M. Markovych, O.~V.
  Viznovych, and M.~V. Tokarchuk.
\newblock Mathematical modeling of subdiffusion impedance in multilayer
  nanostructures.
\newblock \emph{Mathematical Modeling and Computing} \textbf{2}, no.~2, (2015),
  154--159.
\newblock \doi{https://doi.org/10.23939/mmc2015.02.154}.

\bibitem{Kostrobij20184099}
P.~Kostrobij, I.~Grygorchak, F.~Ivashchyshyn, B.~Markovych, O.~Viznovych, and
  M.~Tokarchuk.
\newblock Generalized Electrodiffusion Equation with Fractality of Space–Time:
  Experiment and Theory.
\newblock \emph{The Journal of Physical Chemistry A} \textbf{122}, no.~16,
  (2018), 4099--4110.
\newblock \doi{https://doi.org/10.1021/acs.jpca.8b00188}.

\bibitem{Balescu19954807}
R.~Balescu.
\newblock Anomalous transport in turbulent plasmas and continuous time random
  walks.
\newblock \emph{Phys. Rev. E} \textbf{51}, no.~5, (1995), 4807--4822.
\newblock \doi{https://doi.org/10.1103/PhysRevE.51.4807}.

\bibitem{Tribeche2011103702}
M.~Tribeche and P.~K. Shukla.
\newblock Charging of a dust particle in a plasma with a non extensive electron
  distribution function.
\newblock \emph{Physics of Plasmas} \textbf{18}, no.~10, (2011), 103702.
\newblock \doi{http://dx.doi.org/10.1063/1.3641967}.

\bibitem{Gong2012023704}
J.~Gong and J.~Du.
\newblock Dust charging processes in the nonequilibrium dusty plasma with
  nonextensive power-law distribution.
\newblock \emph{Physics of Plasmas} \textbf{19}, no.~2, (2012), 023704.
\newblock \doi{http://dx.doi.org/10.1063/1.3682051}.

\bibitem{Carreras20015096}
B.~A. Carreras, V.~E. Lynch, and G.~M. Zaslavsky.
\newblock Anomalous diffusion and exit time distribution of particle tracers in
  plasma turbulence model.
\newblock \emph{Physics of Plasmas} \textbf{8}, no.~12, (2001), 5096--5103.
\newblock \doi{http://dx.doi.org/10.1063/1.1416180}.

\bibitem{Tarasov2005082106}
V.~E. Tarasov.
\newblock Electromagnetic field of fractal distribution of charged particles.
\newblock \emph{Physics of Plasmas} \textbf{12}, no.~8, (2005), 082106.
\newblock \doi{http://dx.doi.org/10.1063/1.1994787}.

\bibitem{Tarasov2006052107}
V.~E. Tarasov.
\newblock Magnetohydrodynamics of fractal media.
\newblock \emph{Physics of Plasmas} \textbf{13}, no.~5, (2006), 052107.
\newblock \doi{http://dx.doi.org/10.1063/1.2197801}.

\bibitem{Monin1955256}
A.~S. Monin.
\newblock Uravnenija turbulentnoj difuzii.
\newblock \emph{DAN SSSR, ser. geofiz.} \textbf{2}, (1955), 256--259.

\bibitem{Klimontovich2002}
J.~L. Klimontovich.
\newblock \emph{Vvedenie v fiziku otkrytyh sistem}.
\newblock Moskva Janus (2002).

\bibitem{Zaslavsky2002461}
G.~M. Zaslavsky.
\newblock Chaos, fractional kinetics, and anomalous transport.
\newblock \emph{Physics Reports} \textbf{371}, no.~6, (2002), 461--580.
\newblock \doi{http://dx.doi.org/10.1016/S0370-1573(02)00331-9}.

\bibitem{Tarasov2010}
V.~E. Tarasov.
\newblock \emph{Fractional Dynamics: Applications of Fractional Calculus to
  Dynamics of Particles, Fields and Media}.
\newblock Nonlinear Physical Science. Springer Berlin Heidelberg, 1st edition
  (2010).

\bibitem{Zaslavsky1994110}
G.~M. Zaslavsky.
\newblock {Fractional kinetic equation for Hamiltonian chaos}.
\newblock \emph{Physica D: Nonlinear Phenomena} \textbf{76}, no.~1, (1994),
  110--122.
\newblock \doi{http://dx.doi.org/10.1016/0167-2789(94)90254-2}.

\bibitem{Saichev1997753}
A.~I. Saichev and G.~M. Zaslavsky.
\newblock Fractional kinetic equations: solutions and applications.
\newblock \emph{Chaos} \textbf{7}, no.~4, (1997), 753--764.
\newblock \doi{http://dx.doi.org/10.1063/1.166272}.

\bibitem{Zaslavsky2004128}
G.~M. Zaslavsky and M.~A. Edelman.
\newblock Fractional kinetics: from pseudochaotic dynamics to {Maxwell’s}
  Demon.
\newblock \emph{Physica D: Nonlinear Phenomena} \textbf{193}, no. 1–-4, (2004),
  128--147.
\newblock \doi{http://dx.doi.org/10.1016/j.physd.2004.01.014}.

\bibitem{Nigmatullin2006282}
R.~Nigmatullin.
\newblock `{Fractional}' kinetic equations and ‘universal’ decoupling of a
  memory function in mesoscale region.
\newblock \emph{Physica A: Statistical Mechanics and its Applications}
  \textbf{363}, no.~2, (2006), 282--298.
\newblock \doi{http://dx.doi.org/10.1016/j.physa.2005.08.033}.

\bibitem{Chechkin200278}
A.~V. Chechkin, V.~Y. Gonchar, and M.~Szyd{\l}owski.
\newblock Fractional kinetics for relaxation and superdiffusion in a magnetic
  field.
\newblock \emph{Physics of Plasmas} \textbf{9}, no.~1, (2002), 78--88.
\newblock \doi{http://dx.doi.org/10.1063/1.1421617}.

\bibitem{Gafiychuk2007055201}
V.~V. Gafiychuk and B.~Y. Datsko.
\newblock Stability analysis and oscillatory structures in time-fractional
  reaction-diffusion systems.
\newblock \emph{Phys. Rev. E} \textbf{75}, no.~5, (2007), 055201.
\newblock \doi{https://doi.org/10.1103/PhysRevE.75.055201}.

\bibitem{Datsko2018237}
B.~Datsko and V.~Gafiychuk.
\newblock Complex spatio-temporal solutions in fractional reaction-diffusion
  systems near a bifurcation point.
\newblock \emph{Fractional Calculus and Applied Analysis} \textbf{21}, no.~1,
  (2018), 237--253.
\newblock \doi{https://doi.org/10.1515/fca-2018-0015}.

\bibitem{Kosztolowicz2008066103}
T.~Koszto\l{}owicz and K.~D. Lewandowska.
\newblock Time evolution of the reaction front in a subdiffusive system.
\newblock \emph{Phys. Rev. E} \textbf{78}, no.~6, (2008), 066103.
\newblock \doi{https://doi.org/10.1103/PhysRevE.78.066103}.

\bibitem{Shkilev20131066}
V.~P. Shkilev.
\newblock Subdiffusion of mixed origin with chemical reactions.
\newblock \emph{Journal of Experimental and Theoretical Physics} \textbf{117},
  no.~6, (2013), 1066--1070.
\newblock \doi{https://doi.org/10.1134/S1063776113140045}.

\bibitem{Baron2019052124}
J.~W. Baron and T.~Galla.
\newblock Stochastic fluctuations and quasipattern formation in
  reaction-diffusion systems with anomalous transport.
\newblock \emph{Phys. Rev. E} \textbf{99}, no.~5, (2019), 052124.
\newblock \doi{https://doi.org/10.1103/PhysRevE.99.052124}.

\bibitem{Laskin2000780}
N.~Laskin.
\newblock Fractals and quantum mechanics.
\newblock \emph{Chaos} \textbf{10}, no.~4, (2000), 780--790.
\newblock \doi{http://dx.doi.org/10.1063/1.1050284}.

\bibitem{Laskin20003135}
N.~Laskin.
\newblock Fractional quantum mechanics.
\newblock \emph{Phys. Rev. E} \textbf{62}, no.~3, (2000), 3135--3145.
\newblock \doi{https://doi.org/10.1103/PhysRevE.62.3135}.

\bibitem{Laskin2000298}
N.~Laskin.
\newblock {Fractional quantum mechanics and L\'{e}vy path integrals}.
\newblock \emph{Physics Letters A} \textbf{268}, no. 4-–6, (2000), 298--305.
\newblock \doi{http://dx.doi.org/10.1016/S0375-9601(00)00201-2}.

\bibitem{Laskin2002056108}
N.~Laskin.
\newblock {Fractional Schr\"odinger equation}.
\newblock \emph{Phys. Rev. E} \textbf{66}, no.~5, (2002), 056108.
\newblock \doi{https://doi.org/10.1103/PhysRevE.66.056108}.

\bibitem{Naber20043339}
M.~Naber.
\newblock {Time fractional Schr\"{o}dinger equation}.
\newblock \emph{Journal of Mathematical Physics} \textbf{45}, no.~8, (2004),
  3339--3352.
\newblock \doi{http://dx.doi.org/10.1063/1.1769611}.

\bibitem{Magin20101586}
R.~L. Magin.
\newblock Fractional calculus models of complex dynamics in biological tissues.
\newblock \emph{Computers \& Mathematics with Applications} \textbf{59}, no.~5,
  (2010), 1586--1593.
\newblock \doi{https://doi.org/10.1016/j.camwa.2009.08.039}.

\bibitem{Keshavarz2017663}
B.~Keshavarz, T.~Divoux, S.~Manneville, and G.~H. McKinley.
\newblock Nonlinear Viscoelasticity and Generalized Failure Criterion for
  Polymer Gels.
\newblock \emph{ACS Macro Letters} \textbf{6}, no.~7, (2017), 663--667.
\newblock \doi{https://doi.org/10.1021/acsmacrolett.7b00213}.

\bibitem{Bonfanti20206002}
A.~Bonfanti, J.~L. Kaplan, G.~Charras, and A.~Kabla.
\newblock Fractional viscoelastic models for power-law materials.
\newblock \emph{Soft Matter} \textbf{16}, no.~26, (2020), 6002--6020.
\newblock \doi{http://dx.doi.org/10.1039/D0SM00354A}.

\bibitem{Makris2020}
N.~Makris and E.~Efthymiou.
\newblock Time-Response Functions of Fractional-Derivative Rheological Models
  (2020).

\bibitem{Hobbie2007}
R.~K. Hobbie and B.~J. Roth.
\newblock \emph{{Intermediate Physics for Medicine and Biology}}.
\newblock Springer-Verlag, New York, 4 edition (2007).

\bibitem{Jeon2012188103}
J.-H. Jeon, H.~M.-S. Monne, M.~Javanainen, and R.~Metzler.
\newblock Anomalous Diffusion of Phospholipids and Cholesterols in a Lipid
  Bilayer and its Origins.
\newblock \emph{Phys. Rev. Lett.} \textbf{109}, no.~18, (2012), 188103.
\newblock \doi{https://doi.org/10.1103/PhysRevLett.109.188103}.

\bibitem{Hofling2013046602}
F.~H\"{o}fling and T.~Franosch.
\newblock Anomalous transport in the crowded world of biological cells.
\newblock \emph{Reports on Progress in Physics} \textbf{76}, no.~4, (2013),
  046602.
\newblock \doi{https://doi.org/10.1088/0034-4885/76/4/046602}.

\bibitem{Uchaikin20131074}
V.~V. Uchaikin.
\newblock Fractional phenomenology of cosmic ray anomalous diffusion.
\newblock \emph{Physics-Uspekhi} \textbf{56}, no.~11, (2013), 1074--1119.
\newblock \doi{https://doi.org/10.3367/UFNe.0183.201311b.1175}.

\bibitem{Szymanski2009038102}
J.~Szymanski and M.~Weiss.
\newblock Elucidating the Origin of Anomalous Diffusion in Crowded Fluids.
\newblock \emph{Phys. Rev. Lett.} \textbf{103}, no.~3, (2009), 038102.
\newblock \doi{https://doi.org/10.1103/PhysRevLett.103.038102}.

\bibitem{O'Shaughnessy1985455}
B.~O'Shaughnessy and I.~Procaccia.
\newblock {Analytical Solutions for Diffusion on Fractal Objects}.
\newblock \emph{Phys. Rev. Lett.} \textbf{54}, no.~5, (1985), 455--458.
\newblock \doi{https://doi.org/10.1103/PhysRevLett.54.455}.

\bibitem{Metzler1999431}
R.~Metzler, E.~Barkai, and J.~Klafter.
\newblock {Deriving fractional Fokker-Planck equations from a generalised
  master equation}.
\newblock \emph{Europhysics Letters} \textbf{46}, no.~4, (1999), 431--436.
\newblock \doi{https://doi.org/10.1209/epl/i1999-00279-7}.

\bibitem{Essex2000299}
C.~Essex, C.~Schulzky, A.~Franz, and K.~H. Hoffmann.
\newblock {Tsallis and R\'{e}nyi entropies in fractional diffusion and entropy
  production}.
\newblock \emph{Physica A: Statistical Mechanics and its Applications}
  \textbf{284}, no. 1--4, (2000), 299--308.
\newblock \doi{https://doi.org/10.1016/S0378-4371(00)00174-6}.

\bibitem{Tsallis2001}
C.~Tsallis.
\newblock \emph{Nonextensive Statistical Mechanics and Its Applications}.
\newblock Lecture Notes in Physics 560. Springer-Verlag, Berlin, Heidelberg
  (2001).

\bibitem{Gell-Mann2004}
M.~Gell-Mann and C.~Tsallis.
\newblock \emph{Nonextensive entropy: Interdisciplinary applications}.
\newblock Santa Fe Institute Studies on the Sciences of Complexity. Oxford
  University Press, USA (2004).

\bibitem{Vasconcellos20064821}
A.~R. Vasconcellos, J.~Galv\~{a}o Ramos, A.~Gorenstein, M.~U. Kleinke, T.~G.
  Souza~Cruz, and R.~Luzzi.
\newblock {Statistical Approach To Non-Fickian Diffusion}.
\newblock \emph{International Journal of Modern Physics B} \textbf{20}, no.~28,
  (2006), 4821--4841.
\newblock \doi{https://doi.org/10.1142/S0217979206035667}.

\bibitem{Uchaikin2003810}
V.~V. Uchaikin.
\newblock Anomalous diffusion and fractional stable distributions.
\newblock \emph{Journal of Experimental and Theoretical Physics} \textbf{97},
  no.~4, (2003), 810--825.
\newblock \doi{https://doi.org/10.1134/1.1625072}.

\bibitem{Stanislavsky2004418}
A.~A. Stanislavsky.
\newblock {Probability Interpretation of the Integral of Fractional Order}.
\newblock \emph{Theoretical and Mathematical Physics} \textbf{138}, no.~3,
  (2004), 418--431.
\newblock \doi{https://doi.org/10.1023/B:TAMP.0000018457.70786.36}.

\bibitem{Tarasov2004123}
V.~E. Tarasov.
\newblock {Fractional generalization of Liouville equations}.
\newblock \emph{Chaos} \textbf{14}, no.~1, (2004), 123--127.
\newblock \doi{http://dx.doi.org/10.1063/1.1633491}.

\bibitem{Tarasov200517}
V.~E. Tarasov.
\newblock {Fractional Liouville and BBGKI equations}.
\newblock \emph{Journal of Physics: Conference Series} \textbf{7}, no.~1,
  (2005), 17--33.
\newblock \doi{https://doi.org/10.1088/1742-6596/7/1/002}.

\bibitem{Tarasov2005011102}
V.~E. Tarasov.
\newblock Fractional systems and fractional {Bogoliubov} hierarchy equations.
\newblock \emph{Phys. Rev. E} \textbf{71}, no.~1, (2005), 011102.
\newblock \doi{https://doi.org/10.1103/PhysRevE.71.011102}.

\bibitem{Tarasov2006033108}
V.~E. Tarasov.
\newblock Fractional statistical mechanics.
\newblock \emph{Chaos} \textbf{16}, no.~3, (2006), 033108.
\newblock \doi{http://dx.doi.org/10.1063/1.2219701}.

\bibitem{Tarasov2006341}
V.~E. Tarasov.
\newblock {Transport equations from Liouville equations for fractional
  systems}.
\newblock \emph{International Journal of Modern Physics B} \textbf{20}, no.~3,
  (2006), 341--353.
\newblock \doi{https://doi.org/10.1142/S0217979206033267}.

\bibitem{Tarasov2013663}
V.~E. Tarasov.
\newblock Fractional diffusion equations for open quantum system.
\newblock \emph{Nonlinear Dynamics} \textbf{71}, no.~4, (2013), 663--670.
\newblock \doi{https://doi.org/10.1007/s11071-012-0498-8}.

\bibitem{Tarasov20082984}
V.~E. Tarasov.
\newblock {Fractional Heisenberg equation}.
\newblock \emph{Physics Letters A} \textbf{372}, no.~17, (2008), 2984--2988.
\newblock \doi{http://dx.doi.org/10.1016/j.physleta.2008.01.037}.

\bibitem{Tarasov2005286}
V.~E. Tarasov.
\newblock Fractional hydrodynamic equations for fractal media.
\newblock \emph{Annals of Physics} \textbf{318}, no.~2, (2005), 286--307.
\newblock \doi{http://dx.doi.org/10.1016/j.aop.2005.01.004}.

\bibitem{Tarasov2007237}
V.~E. Tarasov.
\newblock Liouville and Bogoliubov Equations With Fractional Derivatives.
\newblock \emph{Modern Physics Letters B} \textbf{21}, no.~05, (2007),
  237--248.
\newblock \doi{https://doi.org/10.1142/S0217984907012700}.

\bibitem{Tarasov2007163}
V.~E. Tarasov.
\newblock {The Fractional Chapman–Kolmogorov Equation}.
\newblock \emph{Modern Physics Letters B} \textbf{21}, no.~04, (2007),
  163--174.
\newblock \doi{https://doi.org/10.1142/S0217984907012712}.

\bibitem{Tarasov2009179}
V.~E. Tarasov.
\newblock {Fractional generalization of the quantum Markovian master equation}.
\newblock \emph{Theoretical and Mathematical Physics} \textbf{158}, no.~2,
  (2009), 179--195.
\newblock \doi{https://doi.org/10.1007/s11232-009-0015-5}.

\bibitem{Tarasov20121719}
V.~E. Tarasov.
\newblock {Quantum dissipation from power-law memory}.
\newblock \emph{Annals of Physics} \textbf{327}, no.~6, (2012), 1719--1729.
\newblock \doi{http://dx.doi.org/10.1016/j.aop.2012.02.011}.

\bibitem{Tarasov2013102110}
V.~E. Tarasov.
\newblock {Power-law spatial dispersion from fractional Liouville equation}.
\newblock \emph{Physics of Plasmas} \textbf{20}, no.~10, (2013), 102110.
\newblock \doi{http://dx.doi.org/10.1063/1.4825144}.

\bibitem{Kobelev20000002002}
L.~Y. Kobelev.
\newblock {The Multifractal Time and Irreversibility in Dynamic Systems}
  (2000).

\bibitem{Kobelev2000194}
Y.~L. Kobelev, L.~Y. Kobelev, and E.~P. Romanov.
\newblock Kinetic equations for large systems with fractal structures.
\newblock \emph{Doklady Physics} \textbf{45}, no.~5, (2000), 194--197.
\newblock \doi{https://doi.org/10.1134/1.171740}.

\bibitem{Kobelev2002580}
Y.~L. Kobelev, L.~Y. Kobelev, V.~L. Kobelev, and E.~P. Romanov.
\newblock Description of diffusion in fractal media on the basis of the
  Klimontovich kinetic equation in fractal space.
\newblock \emph{Doklady Physics} \textbf{47}, no.~8, (2002), 580--582.
\newblock \doi{https://doi.org/10.1134/1.1505514}.

\bibitem{Lukashchuk2013740}
S.~Y. Lukashchuk.
\newblock {Time-fractional extensions of the Liouville and Zwanzig equations}.
\newblock \emph{Central European Journal of Physics} \textbf{11}, no.~6,
  (2013), 740--749.
\newblock \doi{https://doi.org/10.2478/s11534-013-0229-x}.

\bibitem{Sandev2018015201}
T.~Sandev, Z.~Tomovski, J.~L.~A. Dubbeldam, and A.~Chechkin.
\newblock Generalized diffusion-wave equation with memory kernel.
\newblock \emph{Journal of Physics A: Mathematical and Theoretical}
  \textbf{52}, no.~1, (2018), 015201.
\newblock \doi{https://doi.org/10.1088/1751-8121/aaefa3}.

\bibitem{Sandev2019}
T.~Sandev, R.~Metzler, and A.~Chechkin.
\newblock Generalised Diffusion and Wave Equations: Recent Advances (2019).

\bibitem{Giusti2018013506}
A.~Giusti.
\newblock Dispersion relations for the time-fractional Cattaneo-Maxwell heat
  equation.
\newblock \emph{Journal of Mathematical Physics} \textbf{59}, no.~1, (2018),
  013506.
\newblock \doi{https://doi.org/10.1063/1.5001555}.

\bibitem{Kostrobij201963}
P.~P. Kostrobij, B.~M. Markovych, O.~V. Viznovych, and M.~V. Tokarchuk.
\newblock {Generalized transport equation with nonlocality of space–time.
  Zubarev’s NSO method}.
\newblock \emph{Physica A: Statistical Mechanics and its Applications}
  \textbf{514}, (2019), 63--70.
\newblock \doi{https://doi.org/10.1016/j.physa.2018.09.051}.

\bibitem{Kostrobij2016093301}
P.~Kostrobij, B.~Markovych, O.~Viznovych, and M.~Tokarchuk.
\newblock {Generalized diffusion equation with fractional derivatives within
  Renyi statistics}.
\newblock \emph{Journal of Mathematical Physics} \textbf{57}, no.~9, (2016),
  093301.
\newblock \doi{https://doi.org/10.1063/1.4962159}.

\bibitem{Kostrobij2016163}
P.~Kostrobij, B.~Markovych, O.~Viznovych, and M.~Tokarchuk.
\newblock Generalized electrodiffusion equation with fractality of space-time.
\newblock \emph{Mathematical Modeling and Computing} \textbf{3}, no.~2, (2016),
  163--172.
\newblock \doi{https://doi.org/10.23939/mmc2016.02.163}.

\bibitem{Glushak201857}
P.~A. Glushak, B.~B. Markiv, and M.~V. Tokarchuk.
\newblock Zubarev’s Nonequilibrium Statistical Operator Method in the
  Generalized Statistics of Multiparticle Systems.
\newblock \emph{Theoretical and Mathematical Physics} \textbf{194}, no.~1,
  (2018), 57--73.
\newblock \doi{https://doi.org/10.1134/S0040577918010051}.

\bibitem{Kostrobij201875}
P.~Kostrobij, B.~Markovych, O.~Viznovych, and M.~Tokarchuk.
\newblock Generalized transport equation with fractality of space-time.
  Zubarev's NSO method.
\newblock \emph{CEUR Workshop Proceedings} \textbf{2300}, (2018), 75--78.

\bibitem{Kostrobij201958}
P.~Kostrobij, B.~Markovych, O.~Viznovych, I.~Zelinska, and M.~Tokarchuk.
\newblock Generalized Cattaneo–Maxwell diffusion equation with fractional
  derivatives. Dispersion relations.
\newblock \emph{Mathematical Modeling and Computing} \textbf{6}, no.~1, (2019),
  58--68.
\newblock \doi{https://doi.org/10.23939/mmc2019.01.058}.

\bibitem{Grygorchak2017185501}
I.~I. Grygorchak, F.~O. Ivashchyshyn, M.~V. Tokarchuk, N.~T. Pokladok, and
  O.~V. Viznovych.
\newblock {Modification of properties of
  GaSe$\langle\beta$-cyclodexterin$\langle$FeSO$_4\rangle\rangle$ Clathrat by
  synthesis in superposed electric and light-wave fields}.
\newblock \emph{Journal of Applied Physics} \textbf{121}, no.~18, (2017),
  185501.
\newblock \doi{https://doi.org/10.1063/1.4983097}.

\bibitem{Kostrobij2019289}
P.~P. Kostrobij, B.~M. Markovych, I.~A. Ryzha, and M.~V. Tokarchuk.
\newblock Generalized kinetic equation with spatio-temporal nonlocality.
\newblock \emph{Mathematical Modeling and Computing} \textbf{6}, no.~2, (2019),
  289--296.
\newblock \doi{https://doi.org/10.23939/mmc2019.02.289}.

\bibitem{Zubarev19811509}
D.~N. Zubarev.
\newblock Modern methods of the statistical theory of nonequilibrium processes.
\newblock \emph{Journal of Soviet Mathematics} \textbf{16}, no.~6, (1981),
  1509--1571.
\newblock \doi{https://doi.org/10.1007/BF01091712}.

\bibitem{Zubarev20021and2}
D.~N. Zubarev, V.~G. Morozov, and G.~R\"{o}pke.
\newblock \emph{Statistical mechanics of nonequilibrium processes}, volume 1,
  2.
\newblock Fizmatlit (2002).

\bibitem{Markiv2011785}
B.~Markiv, R.~Tokarchuk, P.~Kostrobij, and M.~Tokarchuk.
\newblock {Nonequilibrium statistical operator method in Renyi statistics}.
\newblock \emph{Physica A: Statistical Mechanics and its Applications}
  \textbf{390}, no.~5, (2011), 785--791.
\newblock \doi{https://doi.org/10.1016/j.physa.2010.11.009}.

\bibitem{Kostrobij202023003}
P.~P. Kostrobij, B.~M. Markovych, and M.~V. Tokarchuk.
\newblock Generalized diffusion equation with nonlocality of space-time. Memory
  function modelling.
\newblock \emph{Condensed Matter Physics} \textbf{23}, no.~2, (2020),
  23003:1--23003:8.
\newblock \doi{https://doi.org/10.5488/CMP.23.23003}.

\end{thebibliography}


 \bigskip \smallskip

 \it

 \noindent
$^1$ Lviv Polytechnic National University \\
12 S. Bandera Str. \\
Lviv -- 79013, UKRAINE  \\[4pt]
e-mail: bohdan.m.markovych@lpnu.ua
\hfill Received: August 18, 2020 \\[12pt]
$^2$  Institute for Condensed Matter Physics \\
of the National Academy of Sciences of Ukraine\\
1 Svientsitskii Str. \\
Lviv -- 79011, UKRAINE \\[4pt]
e-mail: mtok2010@ukr.net

\end{document}